\documentclass[10pt,letterpaper]{article}
\usepackage{opex3}
\usepackage{color}
\usepackage{threeparttable}

\usepackage[utf8]{inputenc}
\usepackage{graphicx}
\usepackage[normalem]{ulem}
\usepackage{float}

\newcommand\micron{\mbox{$\mu$m}}

\begin{document}

\title{In-focus wavefront sensing using non-redundant mask-induced pupil diversity}
\author{Alexandra Z. Greenbaum$^{1,*}$ \& Anand Sivaramakrishnan$^2$}

\address{$^1$Department of Physics and Astronomy, Johns Hopkins University, Baltimore, MD, 21218, USA\\
$^2$Space Telescope Science Institute, 3700 San Martin Dr., Baltimore, MD 21218, USA}

\email{$^*$agreenba@pha.jhu.edu} 

\begin{abstract}
Wavefront estimation using in-focus image data is critical to many
applications.  This data is invariant to a sign flip with complex conjugation
of the complex amplitude in the pupil,  making for a non-unique solution.
Information from an in-focus image taken through a non-redundant pupil mask
(NRM) can break this ambiguity, enabling the true aberration to be determined.
We demonstrate this by priming a full pupil Gerchberg-Saxton phase retrieval
with NRM fringe phase information.  We apply our method to measure simulated
aberrations on the segmented James Webb Space Telescope (JWST) mirror using
full pupil and NRM data from its Near Infrared Imager and Slitless Spectrograph
(NIRISS).
\end{abstract}


\bibliographystyle{osajnl}
\bibliography{ms}

\section{Introduction}
Intensity data from far-field or in-focus imagery of a point source is often
used to determine the aberration in an optical system.  Sometimes the
aberration is an engineering or calibration quantity, as in the case of the
immediately post-launch \textit{Hubble Space Telescope} (HST), and sometimes it
is interesting of itself, as occurs in lensless X-ray microscopy.  The
Gerchberg-Saxton (GS) algorithm \cite{GS72} iteratively accomplishes an
estimation of phase from image intensity and a knowledge of  the pupil
geometry, and does typically converge, but it can seem to converge to a
spurious solution that is a local minimum, or converge to either of the two
ambiguous solutions that are global minima. The unconstrained GS algorithm can
converge to either the true pupil field $P(x)$ or its complex conjugate whose
argument's sign is reversed, $P^*(-x)$ \cite{1986JOSAA...3.1897F}.  Breaking
the two-fold ambiguity can be accomplished in different ways.  A consideration
of more data taken with added defocus (or other kinds of phase diversity) can
result in obtaining the true aberration \cite{Misell1973,1995ApOpt..34.4951K}.
Extra data in the form of pupil diversity can also be used to break the sign
ambiguity of the phase.  Our approach is included amongst the latter class of
methods.

The GS phase retrieval method iteratively applies the known pupil transmission
constraint in the pupil domain and the measured image intensity constraint in
the image domain (e.g.~\cite{1981JOSA...71.1641F}).  Stepping from one domain
to the other is accomplished by a Fourier transform of pupil plane or image
plane complex amplitudes. In the special case where the aberration is
\textit{only} composed of functions where $P(x) = P^*(-x)$ (such as purely
tip/tilt or coma) the phase can be recovered unambiguously with the
unconstrained GS algorithm In practice the algorithm can converge to a local
minimum rather than the true pupil phase unless the initial guess at the pupil
phase is in some heuristic sense fairly close to the correct value. 

\begin{figure*}[htbp!]
\centering
\includegraphics[scale=0.425]{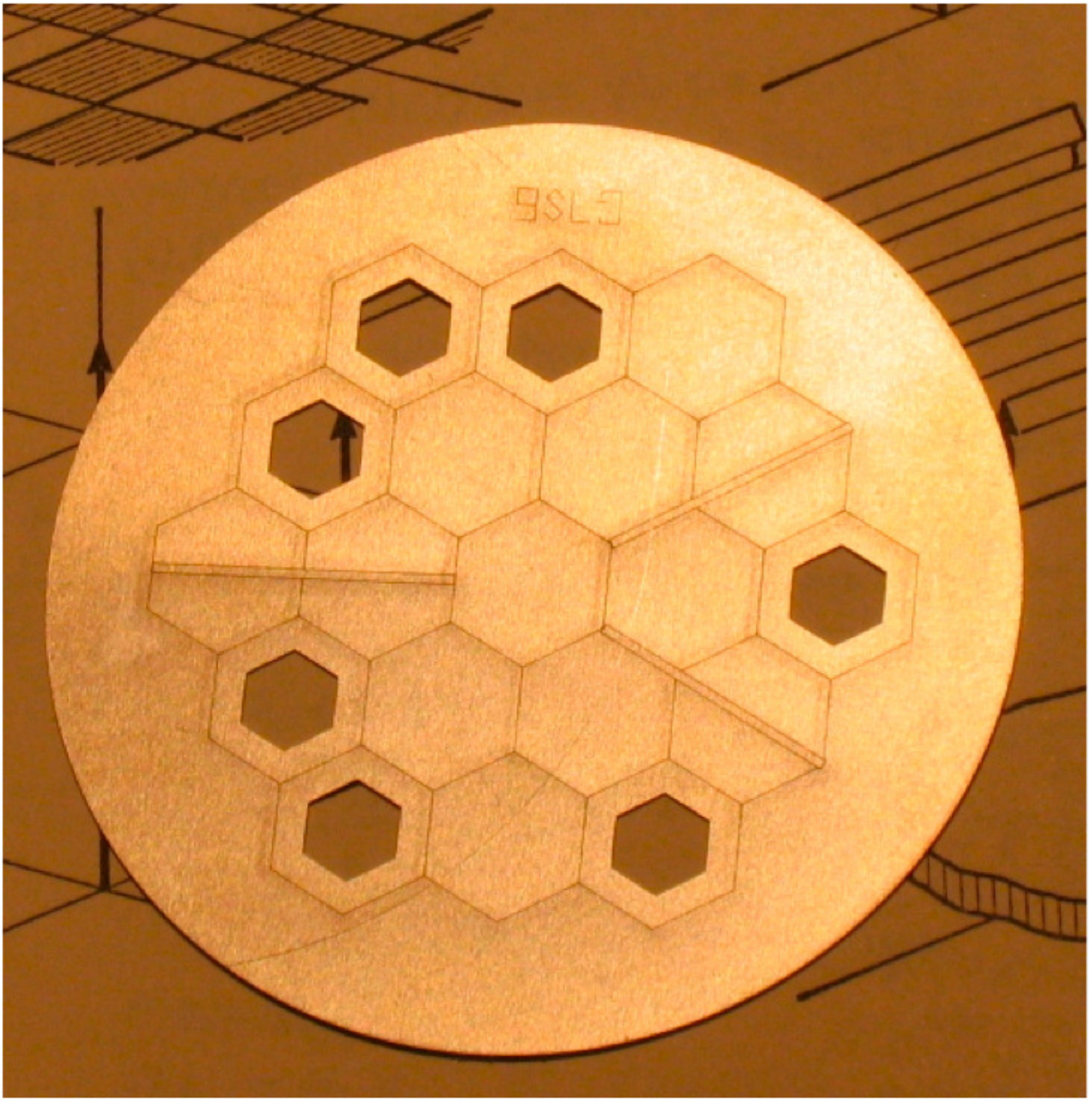}
\includegraphics[scale=0.3]{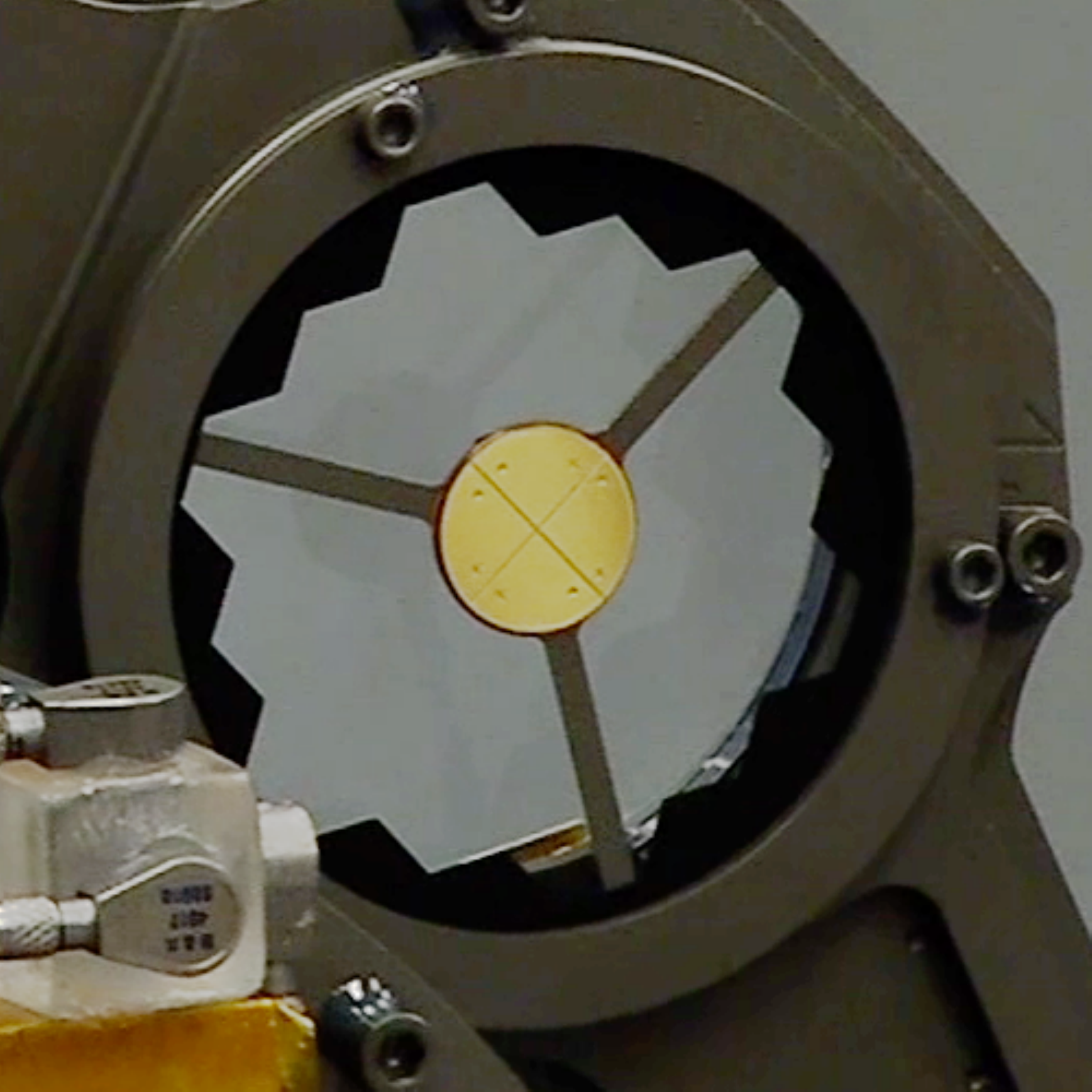}
\caption{JWST NIRISS pupil optics NRM and CLEARP.  A life-sized prototype of
the NRM is shown on the left, and the flight mask CLEARP in the NIRISS Pupil
Wheel on the right.  The JWST primary mirror is reimaged to a 40~mm diameter
pupil in the plane of the NIRISS Pupil Wheel. The high quality of the image of
the primary in NIRISS's entrance pupil \cite{2008SPIE.7010E..3CB}, makes NIRISS
well-suited for our wavefront sensing technique during both commissioning and
routine science operations.}
\label{fig:pupil}
\end{figure*}

\begin{figure*}[htbp!]
\centering
\includegraphics[scale=0.65]{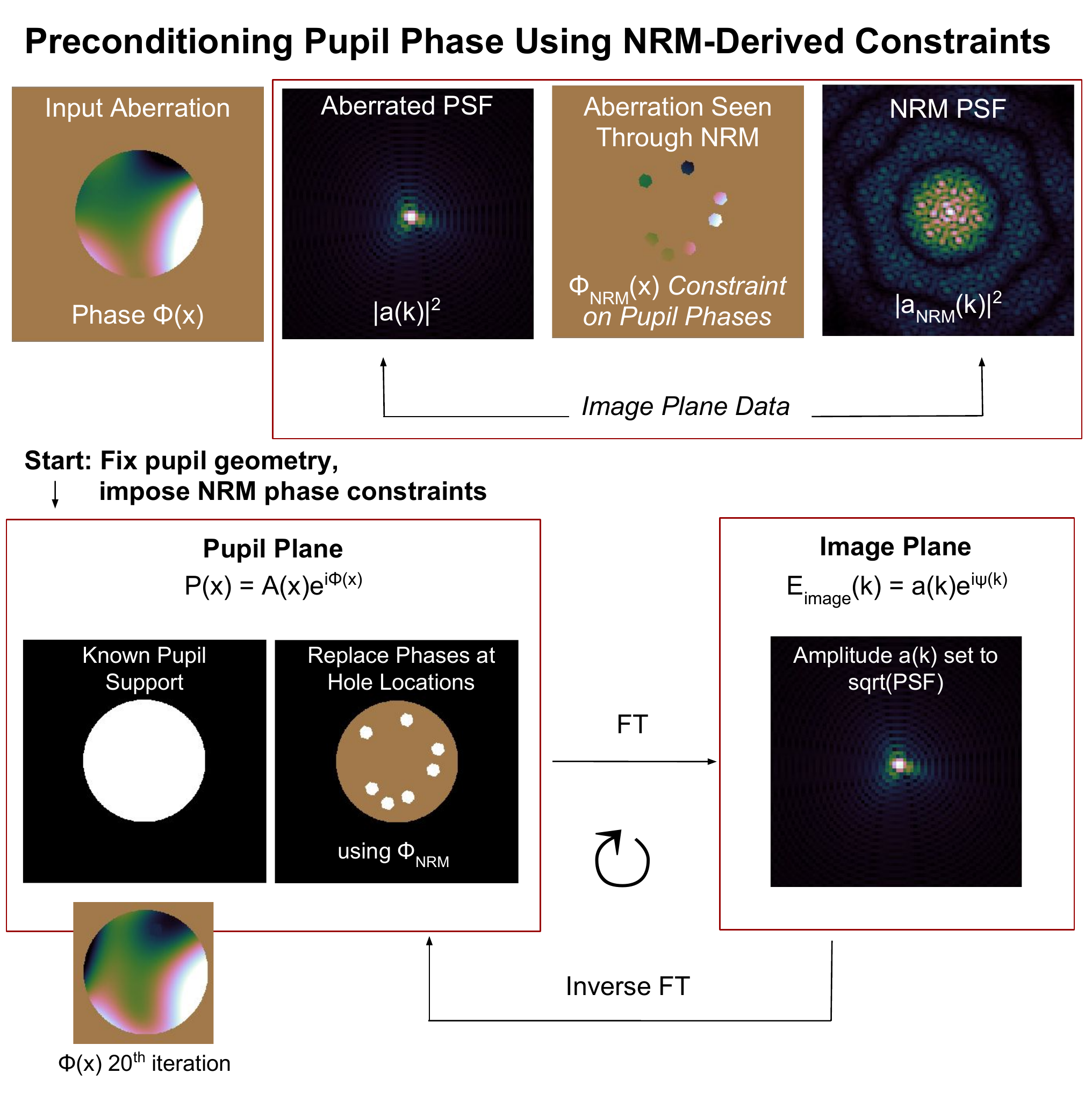}
\caption{An example of true phase aberration, the quantity we are trying to
measure, is shown at the top left. Exposures with the full pupil and the NRM
produce aberrated PSFs. The NRM PSF is used to estimate the phases over each
hole. We assume a known, fixed, pupil support $A(x)$, enforcing the pupil
amplitude to match the known geometry of the pupil in each iteration. In
addition, we also replace phases in the region shown with the average phase
measured over each hole $\phi_{NRM}(x)$, measured from NRM fringe phases
(bottom left two panels). We propagate between pupil and image planes since
$P(x)\rightleftharpoons E_{\mathrm{image}}(k)$. In each iteration we replace
the image field amplitude $a(k)$ at the focus by the square root of the PSF
intensity (bottom right). We show an example of the estimated pupil phase at
the 20th iteration. In this study we remove the NRM constraint after a fixed
number of iterations (typically 100).}
\label{fig:general}
\end{figure*}

We describe a phase retrieval method that uses a pair of in-focus images, one
taken with the full pupil and the other with a non-redundant mask (NRM) in the
pupil.
A NRM consists of a set of holes in a mask where no hole-to-hole vector is
repeated (e.g., Fig. \ref{fig:pupil} (left)).
Our approach can be backup method for fine wavefront sensing that may typically
be required in order to maintain diffraction-limited $2$~$\mu \mathrm{m}$ image
quality on the 6.5~m 18-segment infrared James Webb Space Telescope (JWST),
especially during routine operation and later stages of commissioning.
A flight-ready wavefront sensing scheme using JWST's \textit{Near Infrared
Camera} (NIRCam) has already been developed and tested
\cite{2006SPIE.6265E..0RA}. 
However, JWST’s Near Infrared Imager and Slitless Spectrograph (NIRISS)
\cite{2012SPIE.8442E..2RD} using its two pupil masks, NRM and CLEARP
\cite{2009SPIE.7440E..0YS} (see Fig. 1) can serve as a backup sensing method.
We show that these two pupil masks can be used to measure the telescope’s
aberrations without introducing focus diversity by sweeping the secondary
mirror through focus or placing some of NIRCam's three weak lenses in the beam.

Our backup method reduces mission risk since it provides a second instrument
that can measure JWST's wavefront.
Using both NIRCam and NIRISS also provide wavefront measurements at different
field points, which could assist with secondary mirror alignment when
commissioning JWST.
During routine astronomical observations a pair of images taken with the CLEARP
and NRM pupil masks in one filter can provide a full wavefront measurement to
interferometric accuracy. Such measurements could also support image
deconvolution methods at all wavelengths in NIRISS, where all powered imaging
optics are reflective.  Using in-focus imagery and hardware optimized for
science removes the need for dedicated wavefront sensing hardware, such as weak
defocusing lenses, on future space telescopes.  Our approach does not solve the
persistent problem of non-common path wavefront aberrations in coronagraphs
between the wavefront sensor and the focal plane mask
\cite{2008ApJ...688..701S}, because we measure the wavefront at the science
detector (rather than at the focal plane containing the coronagraphic
occulter).

We quantify the method's performance when faced with realistic limits of noise,
size of the image (number of resolution elements), and wavefront error expected
during certain commissioning phases of JWST. \S\ref{sec:motivation} motivates
our approach and the design choices of our study.
\S\ref{sec:mono}-\ref{sec:poly} show examples of how the algorithm performs
under different conditions using both monochromatic and 8\%~bandwidth images,
matching NIRISS's F480M filter on a continuous circular pupil. In
\S\ref{sec:jwst} we apply the algorithm to a JWST-like pupil with segment tip,
tilt and focus aberrations, as well as global pupil aberrations. We discuss our
results in \S\ref{sec:discussion} in the context of JWST mirror phasing. 

\section{Motivation and Methods}
\label{sec:motivation}

We presume that disjoint segments of the pupil have already been brought to a
common pointing, and to near a common focus.  This places an upper bound on the
sizes of piston and tilt differences over the pupil (given the optical quality
of each segment).  In addition, we assume that every segment in the pupil is
well within the coherence length of the filter bandpass, so that the PSF is
somewhat coherent.  Reaching this coarsely-phased stage is possible using
JWST's NIRCam \cite{2006SPIE.6265E..0RA,
2012OExpr..2029457C,2014OExpr..2212924C} or, as an alternative, with NIRISS and
possibly MIRI \cite{2012OExpr..2029457C,2014OExpr..2212924C}.  After the 18
segments of JWST's primary mirror are coarsely phased we do not expect any
phase wrapping if we only use NIRISS' 4.3~\micron\  or 4.8~\micron\ wavelength
filters.  Our study focuses on the problem of measuring a wavefront aberration
that does not exhibit phase wrapping, in order to explore algorithmic efficacy
rather than address technical complications.

The GS algorithm iterates between pupil plane field $P(x)$ and image plane
field $E_{\mathrm{image}}(k)$, applying constraints in each domain before
returning to the other (by means of a Fourier Transform, for example). For our
image plane constraint we replace the image plane amplitude by the square-root
of the image intensity. We constrain the pupil plane amplitude by forcing it to
be the pupil support function. We modify initial iterations of the GS algorithm
by applying additional constraints in the pupil plane over the regions of the
holes in the NRM.

By analyzing a PSF taken through an $N$ hole NRM we obtain $N(N-1)/2$
\textit{fringe phases}.  Physically, a fringe phase is the piston component of
the optical path delay (OPD) between two holes in the NRM.  We use these fringe
phases to calculate the piston over each hole in the NRM.  This can be done
uniquely by making the $N$ pistons possess a zero mean.

We initialize the pupil phase over the mask holes to the NRM PSF-derived OPDs
and assert zero phase in the pupil everywhere else.  In each subsequent
constrained iteration the pupil phase over the mask holes is replaced with the
NRM-measured phase piston. Elsewhere in the pupil we use the phase provided by
the GS iteration without interpolation. We then smooth the entire pupil phase
by representing it with a small set of low-order polynomials. 

This starts the GS iterations off closer to the true phase, which we show
enables robust and rapid phase retrieval without the need for defocused images.
Our pupil phase estimate approaches the true pupil phase in a just a few
iterations, even when we applied $5\%$ errors to measured piston phases.

During the initial constrained GS iterations the phase over the remainder of
the pupil can drift towards a solution while maintaining agreement with the NRM
image data in selected areas. The basic process is outlined in Fig.
\ref{fig:general}.  After this initial rapid convergence, the NRM-measured
pupil phase constraints are lifted, and the unconstrained GS method converges
to the true phase in most of the cases we tested. Pathological cases where
NRM-derived constraints do not contain any information on the pupil phase can
be constructed (e.g., pure segment tilt with zero hole piston).

We chose to constrain the pupil phase by applying the average hole piston over
the entirety of the mask hole locations rather than over smaller areas (or just
a few points) in the pupil. If the region of replacement is too small it can
add artificial high frequency signal in the reconstructed pupil. In general,
unphysical high frequency (frequencies beyond the limit of the number of
resolution elements in the image) phase will build up during this process,
without smoothing. We smooth the pupil phase each iteration by representing it
with the first 15 Zernike \cite{1976JOSA...66..207N} or Hexike polynomials
\cite{hexikes}, depending on the shape of the pupil. Choosing an optimal set of
basis functions that are better suited to the actual segment geometry
(including obstructions) is beyond the scope of this study. For this study we
add 5\% error to zero mean pupil phases (radians) that would be measured from
the NRM image in order to introduce some measurement error. 

We used 250 pixels across our circular pupils, and up to 1024 pixels for
complicated obstructed apertures.  We have assumed the OPD over each segment
(or the full circular pupil) only contains low spatial frequencies (compared to
the Nyquist frequency of our pupil sampling), so we can represent the pupil
phase over a segment with a few low order polynomials.

\section{Monochromatic Circular Pupil \label{sec:mono}}

In this section we apply our constrained GS phase retrieval algorithm to a
simple circular pupil to demonstrate the principle. Using a known input
wavefront, constructed with a random realization of the first 10 Zernike
polynomials, we simulate images at the NIRISS pixel scale of $64$ mas using a
$6.5$m diameter pupil at $4.8\mu$m to match JWST-NIRISS's F480M filter.  NIRISS
is Nyquist sampled at $\sim$4$\mu$m. Our images are slightly finer than Nyquist
sampled. The technical complications of sub-Nyquist image sampling may be
addressable using existing techniques \cite{1999_fienup}. Our goal is to
measure this input wavefront. We apply our NRM phase constraint for the first
100 iterations and then remove the constraint, allowing the algorithm to run
until it converges to a state where the solution is stable between consecutive
iterations.  We report two relevant quantities, the \textit{true error} and the
\textit{convergence}. The \textit{true error} is the difference between the
actual zero-mean wavefront and the pupil phase estimate. We use the true error
to evaluate the algorithm performance in this study. The \textit{convergence}
is the difference between pupil phase in consecutive iterations. The
convergence criterion sets the stopping point of our iterations.  We set our
numerical convergence criterion to $10^{-6}$ radians difference between
iteration $i$ and $i-1$. We chose this criterion to be smaller than the
expected true error. In a practical situation, the size of the true error will
have to be estimated beforehand with simulations, such as ours, and optical
testing.

We damp each iteration, computing the pupil phase as 80\% of the new solution
and 20\% of the solution from the previous iteration. We did not optimize the
damping in this study. Finally we smooth our estimated pupil phase by
representing it with the first 15 Zernike polynomials when using a circular
pupil. In the case of hexagonal mirror segments we use equivalent 15 Hexike
polynomials on each segment.  The difference between iterations that decides
the convergence condition is calculated inside a slightly undersized pupil to
avoid any edge effects.

\subsection{Concept: Phase Retrieval With and Without Constrained GS (Noiseless Case)}

\begin{figure*}[htbp!]
\centering
\includegraphics[scale=0.65]{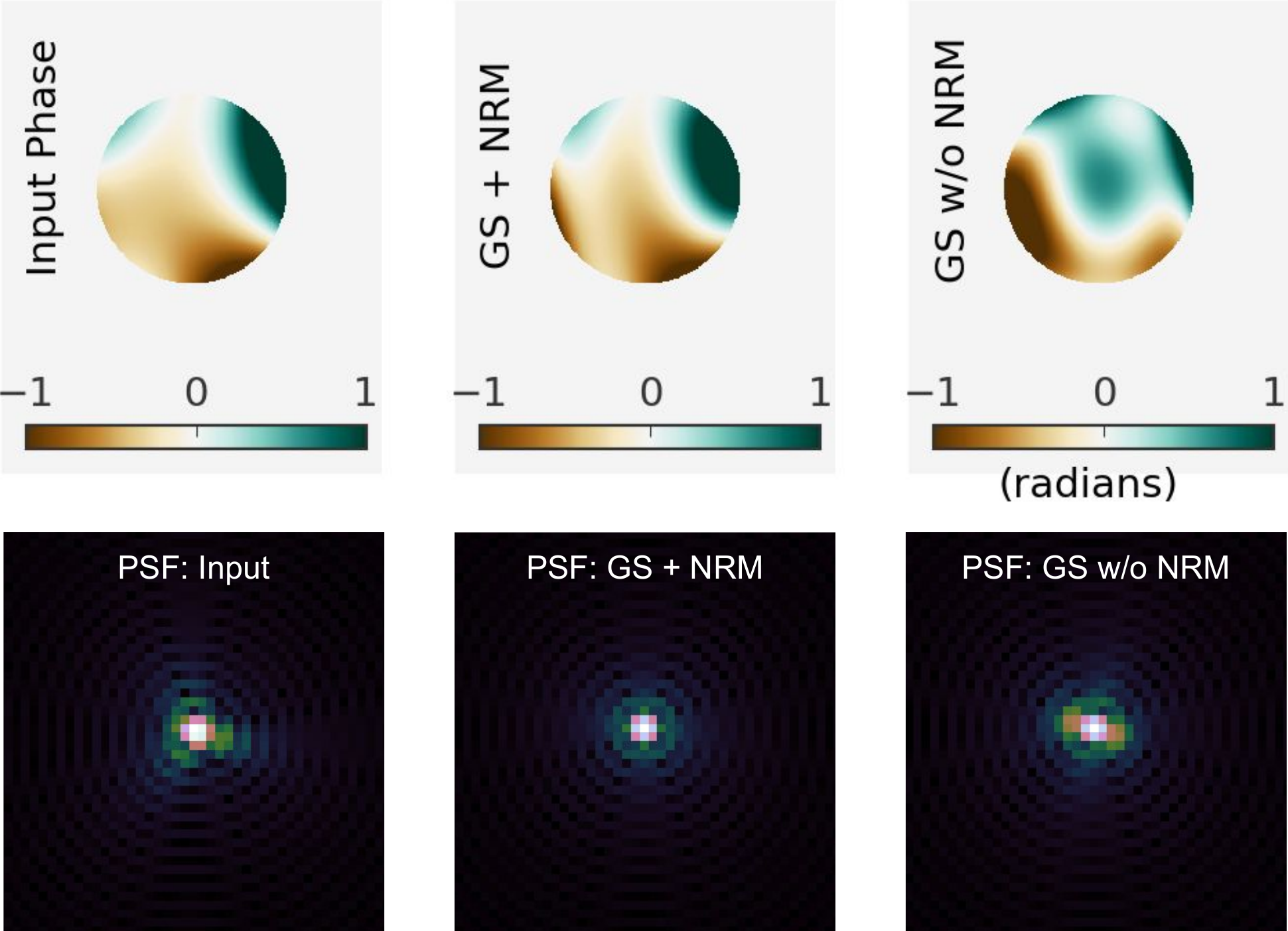}
\caption{\textbf{Top}: Noiseless monochromatic phase retrieval of input
aberration on the leftmost panel with 0.62 radians rms wavefront error. We
display the phase retrieval with NRM constraint (middle panel) and with out
(rightmost panel). With perfect correction the constrained phase retrieval can
sense the wavefront to residual of 0.10 radians rms phase, while the
unconstrained approach produces a residual of 0.53 radians rms phase. At
$\lambda$~=~4.8$\mu$m, 0.1 radians corresponds to $\sim$76 nm of phase.
\textbf{Bottom}: PSFs corresponding the initial PSF in the leftmost panel and
the residual wavefronts in the middle and rightmost panels}
\label{fig:noiseless}
\end{figure*}

\begin{figure*}[htbp!]
\centering
\includegraphics[scale=0.4]{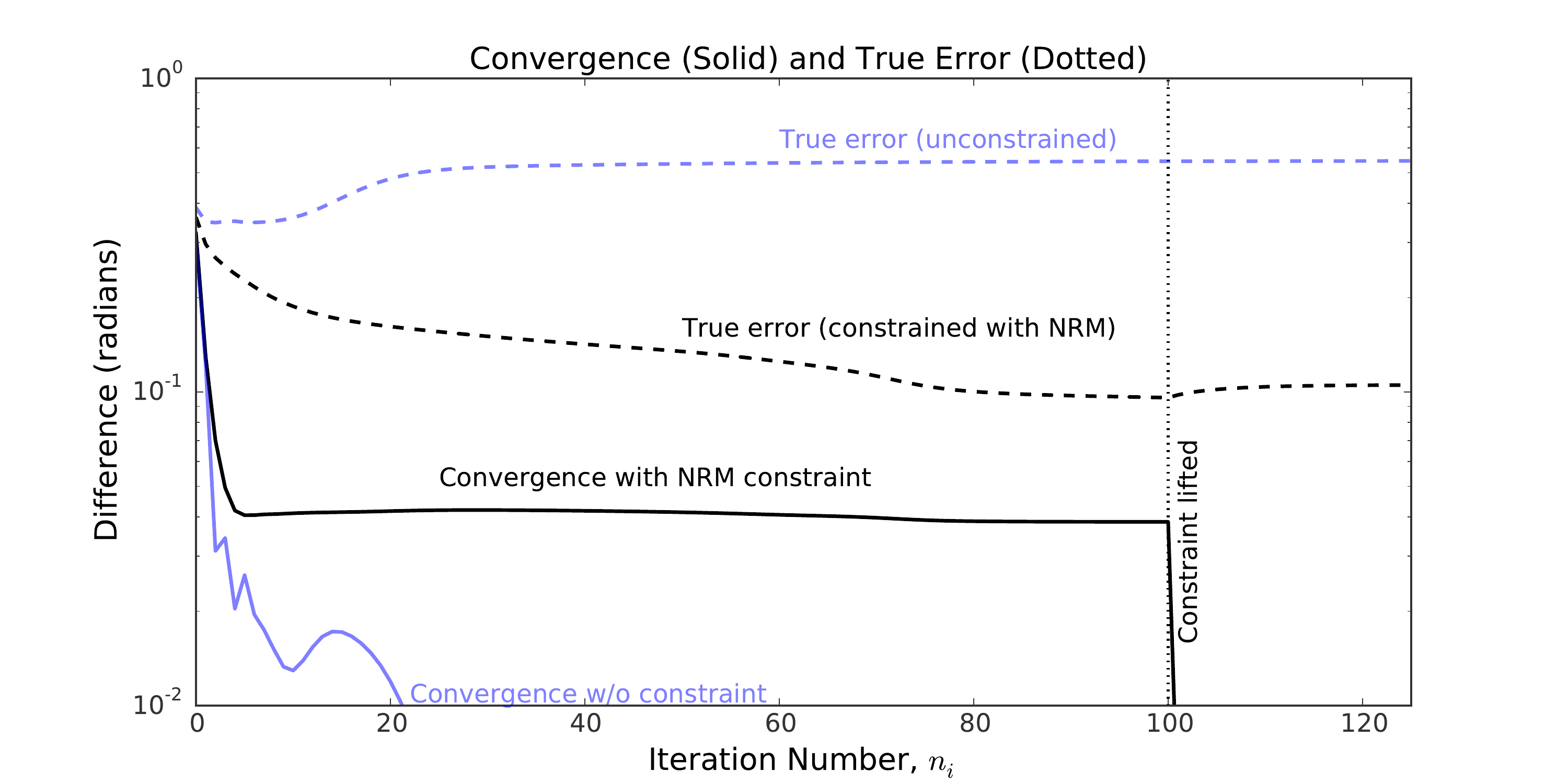}
\caption{Algorithm \textit{convergence}, the difference between pupil phase at
$n_{i}$ and $n_{i-1}$ iterations, is shown in the solid curves. The
\textit{true error}, the difference between the true aberration and the
measured phase, is shown in the dashed curves. These reflect the trials shown
in Fig. \ref{fig:noiseless}. We remove the constraint after 100 iterations
(vertical dotted line) and allow the algorithm to converge. Applying the NRM
constraint resolves the GS degeneracy and allows the algorithm to converge in
the correct solution. In the unconstrained case the algorithm converges quickly
to the wrong solution.}
\label{fig:convergence}
\end{figure*}

We demonstrate the advantage of our constrained approach compared to pure GS
phase retrieval using a noiseless image made from an unobstructed circular
pupil. Figure \ref{fig:noiseless} compares the constrained case (using NRM
fringe phases) with the unconstrained Gerchberg-Saxton algorithm. The initial
pupil has 0.62 radians rms of phase ($\sim460~\mathrm{nm}$). We convert angular
measure to a physical distance assuming a wavelength of $4.8$~$\mu$m, the
central wavelength of the JWST filter we model. Figure \ref{fig:convergence}
displays the true error and convergence for both the constrained and
unconstrained cases.  We refer to the true error as \textit{residual wavefront}
henceforth. In the unconstrained case the algorithm converges quickly (Fig.
\ref{fig:convergence} blue solid line) but to the wrong solution -- the true
error increases over time (blue dotted line). The residual wavefront in this
unconstrained case is 0.53 radians ($\sim$400 nm) rms phase, as seen in the
rightmost panel of Fig. \ref{fig:noiseless}. In the constrained case the
residual wavefront error falls quickly in just a few iterations. The final
residual wavefront using the constrained GS algorithm has 0.1 radians ($\sim$76
nm) (middle panel of Fig. \ref{fig:noiseless}). 

In some cases, after steadily decreasing, the true error between the measured
wavefront and the true aberration, increases a small amount when the constraint
is lifted and the algorithm is allowed to ``relax," as shown in the black
dotted line in Fig. \ref{fig:convergence}. The performance details are most
likely due to our specific choices of damping and smoothing and we anticipate
that improvements can be made in the practical implementation of this approach.
For the remaining sections of this study we focus on the major limiting factors
that will guide the observations necessary to measure the wavefront from
in-focus images. 

\subsection{Image Size Dependence}

\begin{figure*}[htbp!]
\centering
\includegraphics[scale=0.4]{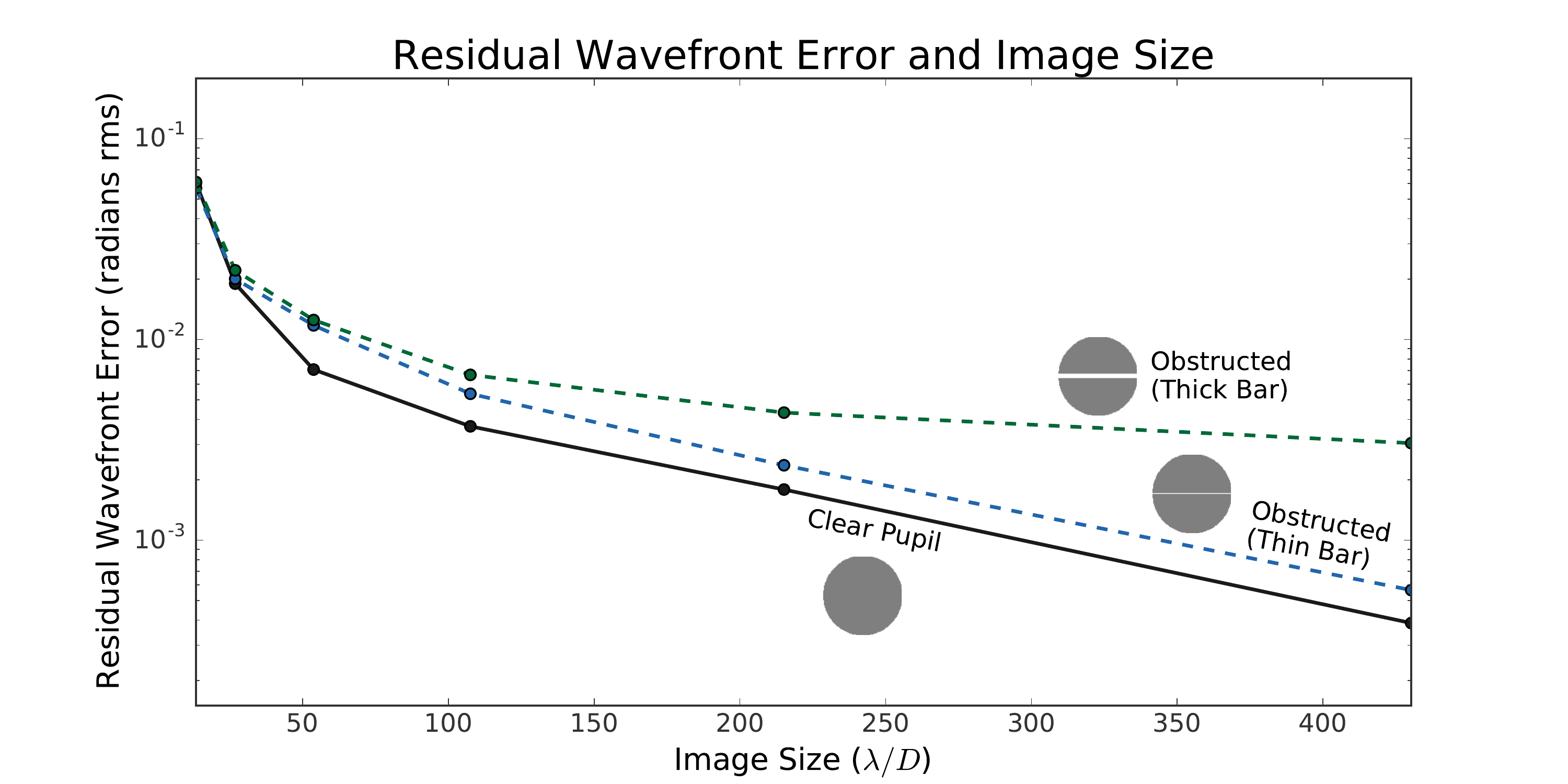}
\caption{We compare reconstruction error with image size (number of resolution
elements in the image). Pupil obstructions introduce more error with increasing
obstruction size. }
\label{fig:fov}
\end{figure*}

The reconstruction performance is ultimately limited by the size of the image
used (in units of the resolution element $\lambda/D$). Roughly speaking, by
limiting the measured spatial frequencies in the pupil, the image size governs
how well we can correct the wavefront. This image size dependence applies only
to the full pupil image, as the NRM image contributes only a guiding first
estimate. In Fig. \ref{fig:fov} we compare the residual error in the measured
wavefront with the number of resolution elements in the PSF. We find that image
sizes of $\sim25\lambda/D$ can be used to correct the wavefront with acceptably
small residual wavefront error. For the remaining simulations in Sections
\ref{sec:mono} and \ref{sec:poly} we used an image size of 128 pixels, which
corresponds to $\sim 50 \lambda/D$. Deep coadded exposures are likely to be
needed for the required dynamic range.  In many of the following cases the
image size sets the floor for our wavefront sensing accuracy.

As described in Perrin et al. (2003), the direct correspondence between radial
distance from the center of the PSF and the spatial frequency in the pupil
plane is only true for Strehl ratios above about 90\%, where the 1st order
expansion of the PSF is a good representation. But for lower Strehl ratios the
correspondence is only an approximation.  Higher order PSF expansion terms show
cause frequency mixing, so phase terms with a spatial frequency of eg. $M$
cycles across the pupil  can place speckles at harmonics located at $2M$, $3M$,
\dots\ resolution elements from the core. Since we use images with Strehl
ratios below 90\%, we find image size limits our wavefront sensing precision.

In the presence of pupil obstructions (a single spider in Fig. \ref{fig:fov})
we still find acceptable, albeit slightly reduced performance; our accuracy
improves with increasing image size. The larger obstruction limits the
reconstruction accuracy most.  We suspect that a different approach to
smoothing each iteration may improve this performance. To mitigate the effects
of pupil obstructions, in Section \ref{sec:jwst}, where we simulate JWST-like
segment gaps and spiders, we use larger image size of 524 pixels, approximately
70 $\lambda/D$.

\subsection{Capture Range}

\begin{figure*}[htbp!]
\centering
\includegraphics[scale=0.4]{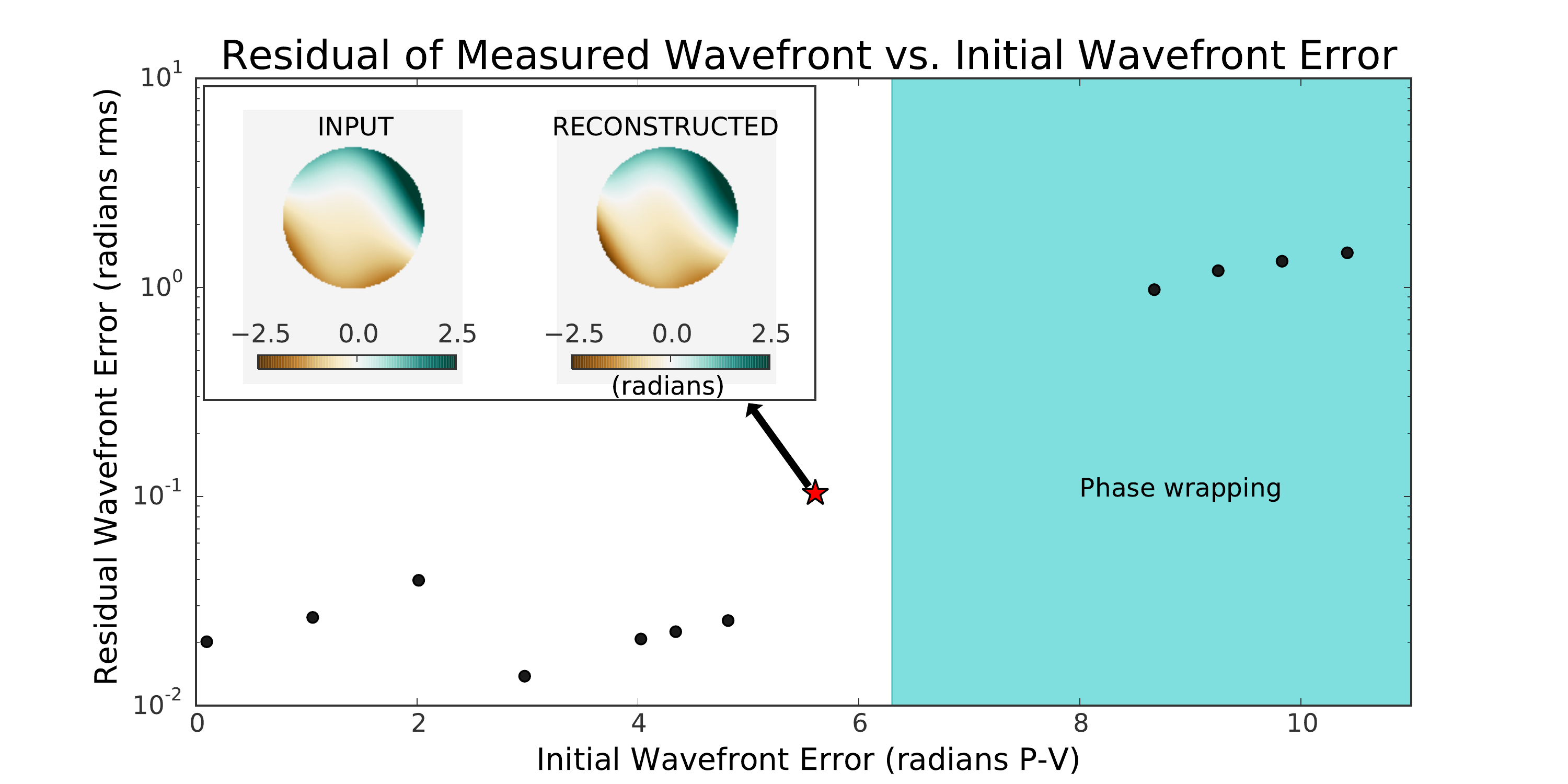}
\caption{The residual wavefront error with increasing peak to valley input
aberration. The inset shows wavefront errors $>\pm2$ radians can be corrected
to $\sim$0.1 radians.  }
\label{fig:wfe}
\end{figure*}

In the noiseless case (using a $50 \lambda/D$ image size) we can recover the
wavefront with a standard deviation $\sigma=10^{-2}$~radians residual error
between the true aberration and the estimated wavefront from our algorithm. Our
input aberrations do not exceed $2\pi$ radians P-V wavefront error because we
do not implement any phase unwrapping in this study. Figure \ref{fig:wfe}
displays the residual between the reconstructed wavefront and the true
aberration as a function of P-V wavefront error of the true aberration,
indicating the region above $2\pi$ radians where phase wrapping occurs.
Incorporating phase unwrapping procedures may improve performance in this
region. 

Other in-pupil approaches to wavefront sensing use an interferometric analysis
with an asymmetric pupil \cite{Martinache_APWFS} or the differential optical
transfer function (dOTF) using a known pupil modification
\cite{2015JATIS...1b9001C}. The dOTF approach is similar to ours in that it
does not need to operate in closed control loop, and it also uses two in-focus
images. However, it requires an additional hardware component or specific
hardware capability to produce a small pupil obstruction. We discuss the
operational differences between our approach and the dOTF method further in
Section \ref{sec:discussion}. Similar to our approach, the asymmetric pupil
wavefront sensor (APWFS) can use existing science hardware in the case of JWST,
but it works iteratively in closed loop. Our monochromatic case, where we have
not applied phase unwrapping (e.g. Fig. \ref{fig:wfe}), has similar capture
range to APWFS.

Multiwavelength NRM data has been used to resolve phase wrapping ambiguity
\cite{2012OExpr..2029457C}. Since phase unwrapping is routinely used in
successful Gerchberg-Saxton methods \cite{1995ApOpt..34.4951K}, combining these
approaches may work well on continuous pupils.  Phase unwrapping on segmented
apertures raises some fundamental issues that are beyond the scope of this
study. Multiwavelength imaging may help resolve some of these issues. The APWFS
works only in the regime where the small angle approximation is valid; when the
wavefront errors are larger than $\sim1$ radians it breaks down \cite{PopeWFS}.
Our approach does not use a small angle approximation. The APWFS operates in a
closed control loop with wavefront correction, whereas our approach provides a
wavefront estimate with just one set of images. We note that the APWFS
measurement could be used to constrain the Gerchberg-Saxton algorithm instead
of pistons derived from NRM fringes, so that our initially constrained approach
can be used with other JWST instruments that do not contain an NRM. 

\subsection{Photon Noise and Exposure Time}

\begin{figure}
\centering
\includegraphics[scale=0.4]{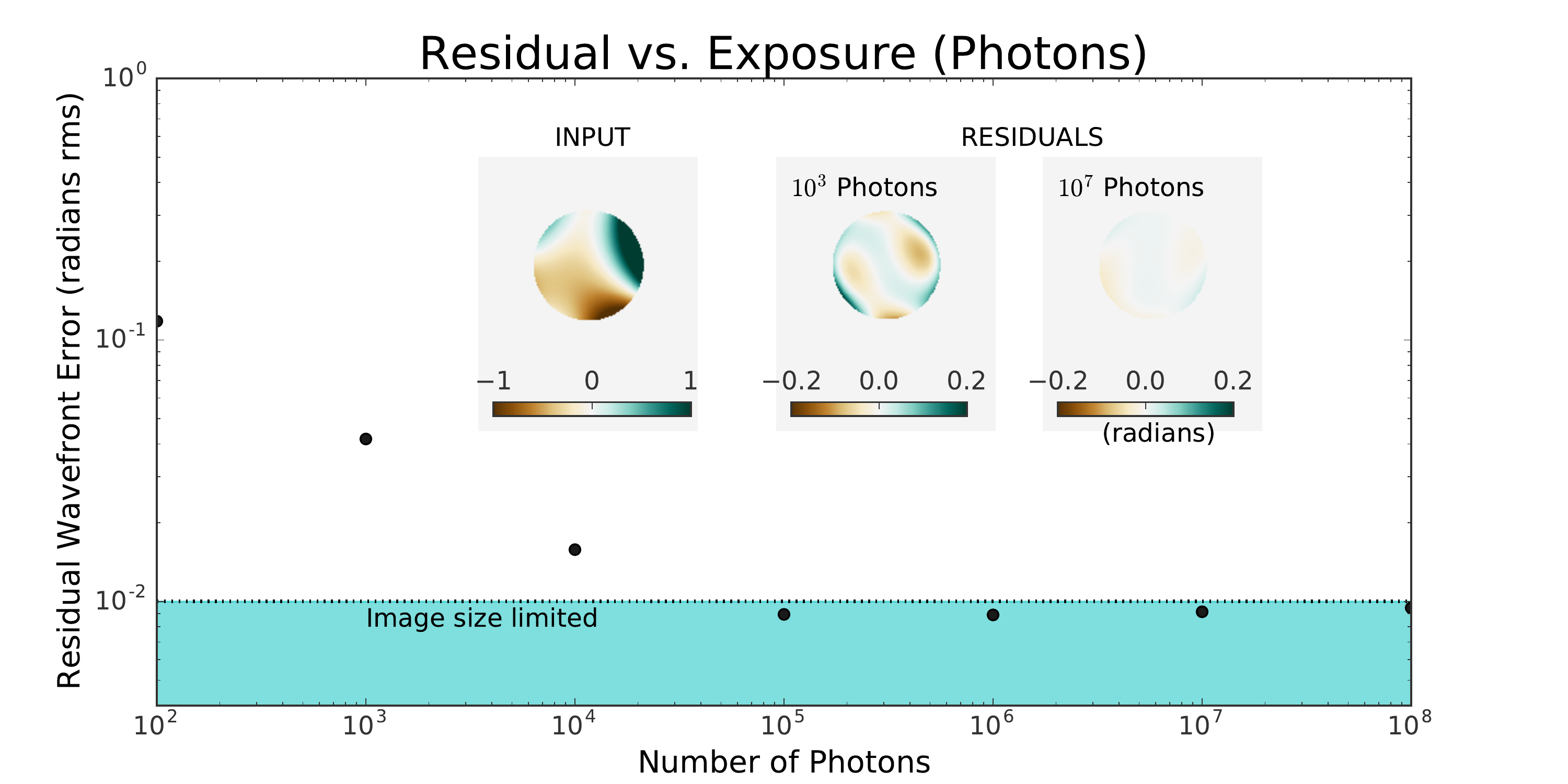}
\caption{We compare results in the case of monochromatic images with photon
noise between 1e3 and 1e7 photons. The leftmost panel shows the input pupil
phase. The middle and left panels are the measured wavefront in the case of
$10^3$ and $10^7$ photons. In practice on NIRISS $10^6$ photons will be easily
acquired on bright stars.}
\label{fig:pnoiseramp}
\end{figure}

Our procedure requires two exposures, one with the full pupil, one with the
NRM. We can tolerate fewer counts in the NRM images since the fringe phases are
only used to move the estimated pupil phase in the right direction during the
first few iterations. Here we consider the necessary exposure time for the full
pupil image in the presence of photon noise. We find that even in shallow full
pupil exposures, we can reconstruct the wavefront to an accuracy of $\sim$0.1
radians rms or better (compared to 0.62 radians rms in the initial aberrated
pupil), as summarized in Fig. \ref{fig:pnoiseramp}. The inset plot shows two
examples at $10^3$ and $10^7$ photons. While the residual error is larger for
an exposure of $10^3$ photons, the wavefront is recovered to $\sim$0.04
radians. Beyond $\sim10^5$ photons we reach the reconstruction limit set by our
50 $\lambda/D$ image size.

Shallow exposure images with the full pupil can be obtained quickly on bright
sources with NIRISS. For example, a 7.5 magnitude star in NIRISS F480M filter
will take of order seconds to reach $10^7$ photons. For the NRM images $10^6$
photons should be more than sufficient to achieve better than the 5\% precision
in pupil phases that we use in these simulations \cite{LGalgo}. It will take
seconds to reach $10^6$ photons on the same 7.5 magnitude star with the NRM. 

\section{Finite Bandwith Images matching NIRISS's F480M Filter \label{sec:poly}}

\begin{figure}
\includegraphics[scale=0.4]{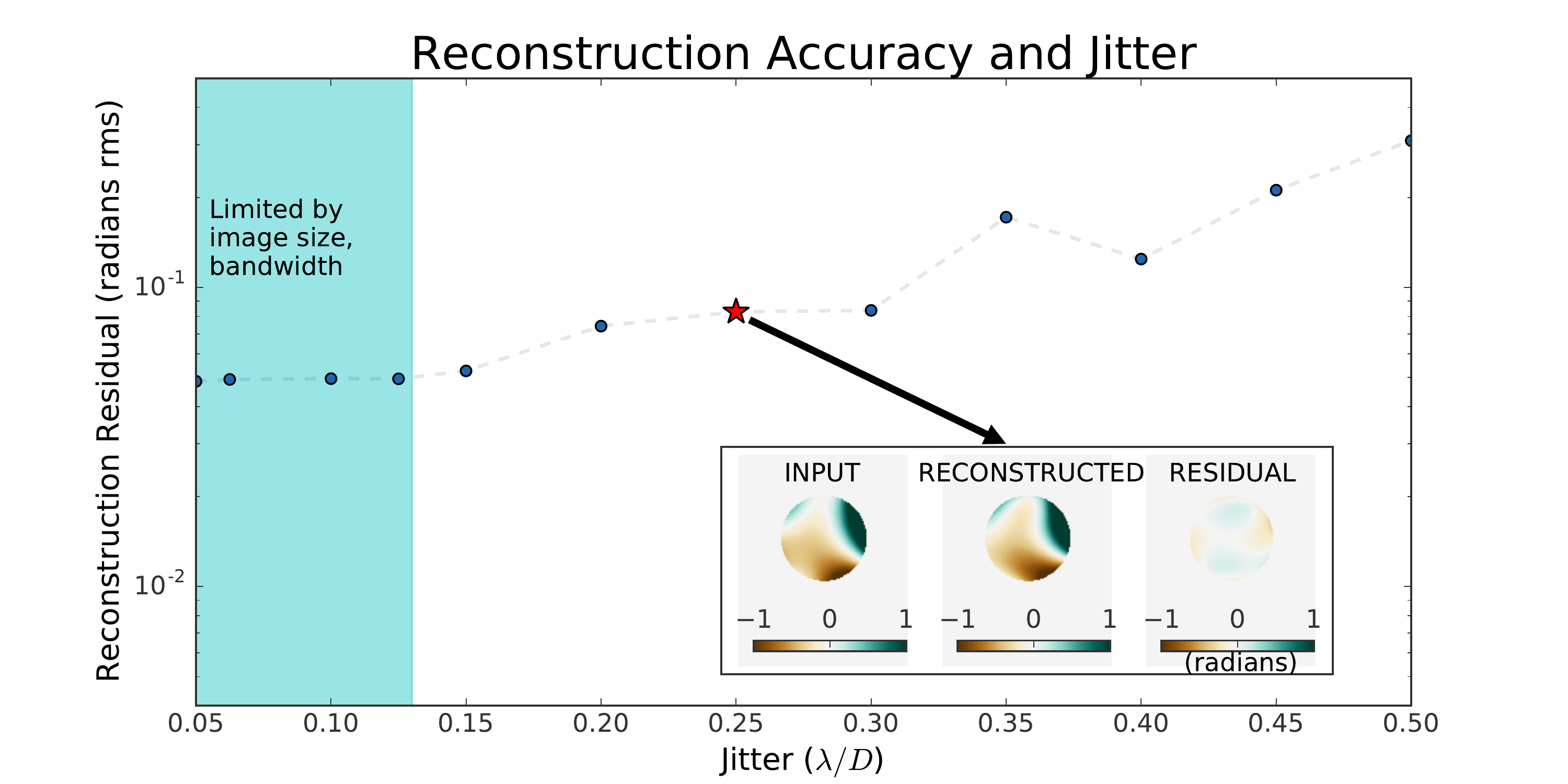}
\caption{We analyze the effect of pointing jitter of up to nearly a full pixel
on the rms residual error for 8\% bandwidth images that have $10^6$ photons. In
this case, below about a quarter pixel of jitter, reconstruction error is
limited by image size, consistent with the $50\ \lambda/D$ point in Fig.
\ref{fig:fov}. The inset plot shows an example of the reconstruction when the
pointing jitter is 0.25 $\lambda/D$. The residual wavefront error in this case
is $<0.1$ radians rms compared to 0.62 radians rms before correction).}
\label{fig:jitterpnoise}
\end{figure}

In reality, images will be chromatic and will contain other sources of noise,
such as from pointing jitter.  In this section we extend our analysis to
polychromatic images matching the bandwidth of NIRISS's 8\% F480M filter. We
consider constant transmission across the filter bandwidth centered at 4.8
$\mu$m. We additionally add pointing jitter by convolving the PSF with a
Gaussian of standard deviation $\sigma  = $ pointing jitter. 

The algorithm reconstructs phase errors in the wavefront at the central
wavelength. The chromatic effects of our 8\% bandwidth images increase the
threshold reconstruction error.  Above $\sim$0.15 $\lambda$/D, jitter is the
larger source of error. We summarize these results in Fig.
\ref{fig:jitterpnoise}, and mark the region where the simulation is limited by
other sources of error (image size and chromatic bandwidth). The inset plot
shows an example of our results in the particular case of $0.25\lambda/D$
jitter and $10^6$ photons (data point indicated by a star marker). In our
example, which has less than ideal conditions, we are able to recover the
wavefront very well, with a residual wavefront error of $<0.1$ radians rms. We
investigated this effect up to about a full pixel of jitter, 0.5 $\lambda/D$,
where performance is degraded, but still produces a residual wavefront error of
$\sim0.3$ radians rms.

The JWST pointing requirement is $7~\mathrm{mas}$ rms
\cite{2006SSRv..123..485G}, which is $\sim$0.05~$\lambda/D$ jitter at
4.8~$\mu\mathrm{m}$. In our illustrative example we find that our algorithm is
limited by the PSF image size (128 pixels) when the jitter is less than
$\sim\lambda/8D$. Up to about a quarter pixel of jitter is correctable to
$\sim$0.05 radians rms (or $\sim$40 nm at the 4.8~$\mu$m wavelength).  If
JWST's pointing jitter remains at or below requirement, it will not be a
limiting factor to this kind of wavefront reconstruction.

\section{Correctable Wavefront Errors on the JWST Segmented Mirror \label{sec:jwst}}

JWST mirror segments have seven actuators.  These control the six solid
body degrees of freedom as well as the segment radius of curvature
\cite{2006SPIE.6265E..0RA}.  Individual JWST mirror segments have been measured
to have very small aberrations, and are designed to be stiff.  The
commissioning plan is to adjust segment radius of curvature only during fine
phasing \cite{2006SPIE.6265E..0XC}, so the range of radius of curvature
actuation need only be of the order of $\sim 100$~nm at ($\sim 0.13 $~radians
at $4.8\mathrm{\mu m}$).  Prior fine phasing, pupil aberrations (including
global defocus) are expected to be dominated by misplaced primary mirror
segments, and, during early commissioning, from a misaligned secondary.

In this section we consider piston, tilt and focus segment aberrations, errors
which are expected from segment drifts between phasing. We apply our algorithm
to images made with a pupil matching the JWST primary hexagonal segments and
mirror obstructions. For this example we do not consider the central
obstruction in the CLEARP pupil (Fig. \ref{fig:pupil} right). We use the Hexike
polynomials implemented in WebbPSF \cite{webbpsf,2014SPIE.9143E..3XP} for
smoothing the measured wavefront each iteration. We do not assume anything
about the particular set of Hexike terms present in the wavefront. In practice,
it may be useful to limit smoothing modes to only piston, tip/tilt, and focus.
In these test cases, we consider images that are sampled three times better
than NIRISS's F480M filter to focus on the algorithm performance for a
segmented pupil rather than sampling effects. We use images of size 524 pixels
($\sim74 \lambda/D$).

\subsection{Segment Tilt\label{jwsttilt}}

In the special case of pure, random segment tip/tilt with no segment piston or
higher order aberrations, constraining the GS with NRM data (which measures
hole piston) only supplies zero piston constraint over all holes. In our pure
segment tilt test case the constrained GS did not recover all the segment tilts
correctly. Because this is a special case, we address segment tilt separate
from piston and defocus.

\begin{figure*}[htbp!]
\centering
\includegraphics[scale=0.8]{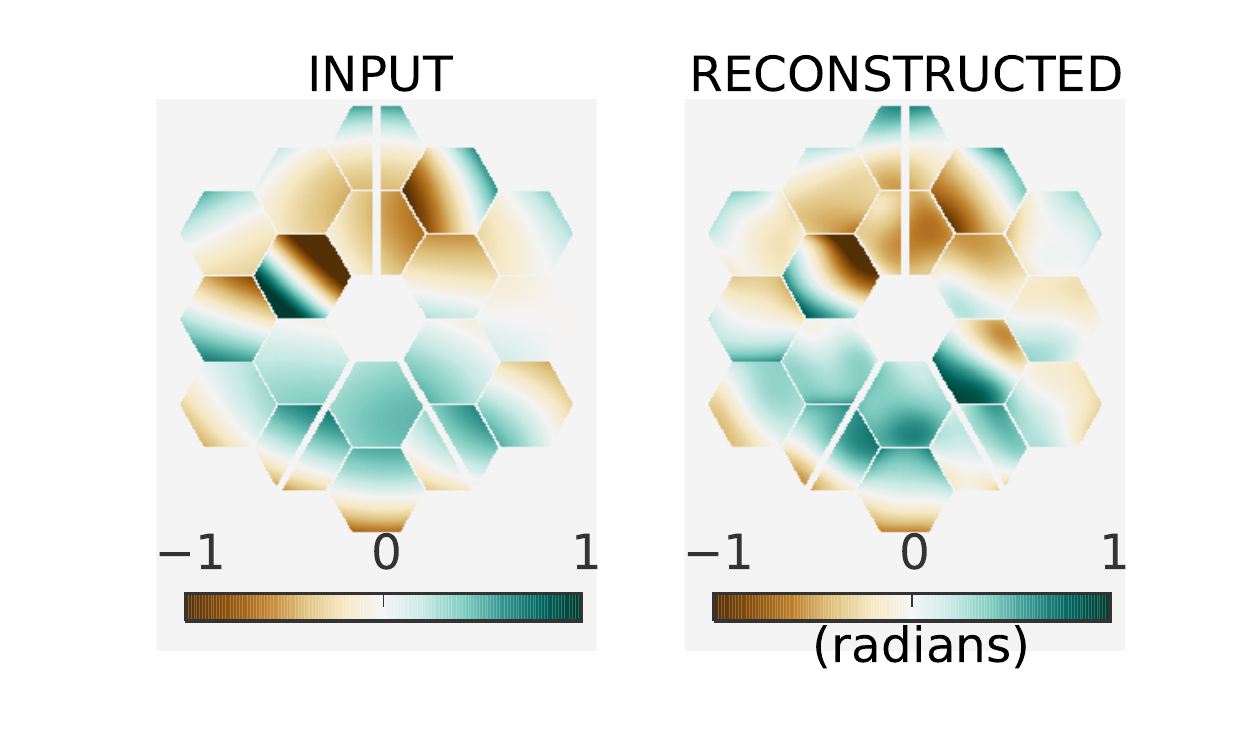}
\caption{Segment tip/tilt is mixed with a global coma.  The initial wavefront
errors are 0.35 radians rms and the reconstruction residuals are 0.14 radians
rms. }
\label{fig:tiltcoma}
\end{figure*}

Once mixed with any other (symmetric) phase, NRM can break the GS ambiguity.
When we added coma (a global aberration that often arises from secondary mirror
misalignment) the wavefront was retrieved unambiguously (Fig.
\ref{fig:tiltcoma}). We also mix tilt with segment piston and focus, and again
recover the wavefront (Fig. \ref{fig:jwst}).

\subsection{Combinations of Segment Piston, Tilt and Focus\label{sec:pttf}}

\begin{figure*}[htbp!]
\centering
\includegraphics[scale=0.75]{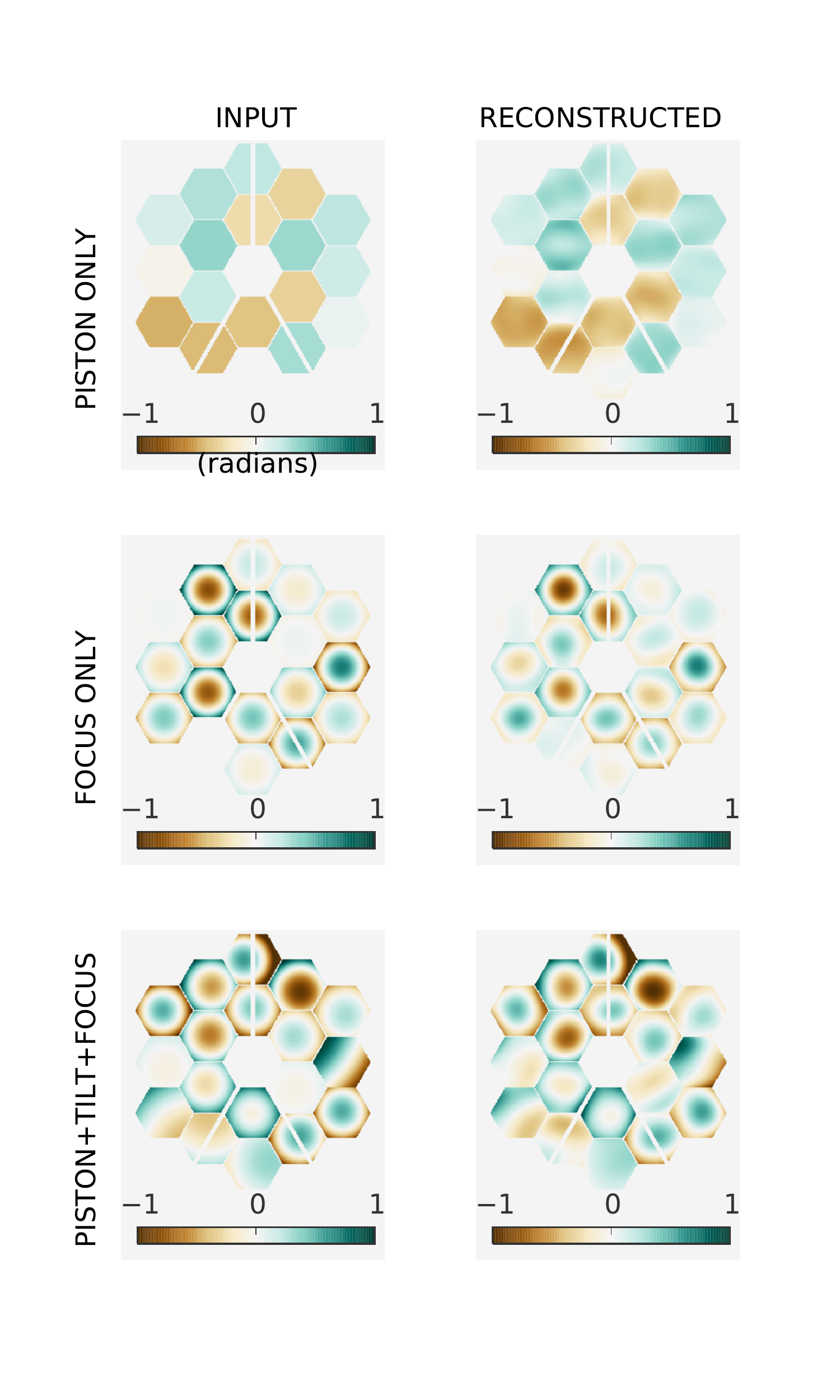}
\caption{We simulate segment piston, tip/tilt, and focus aberrations on the
JWST pupil simulating the F480M filter bandwidth and $\sim7$~mas jitter. The
left column shows the input and the right the reconstructed wavefront. In the
case of pure segment piston (top) the input wavefront has 0.27 radians rms
global wavefront error and the reconstruction residual has 0.06 radians rms.
For the case of pure segment defocus (middle) the input wavefront produces a
global wavefront error of 0.25 radians rms, while the reconstructed wavefront
has residual error of 0.1 radians rms. In the combined piston, tip/tilt, and
focus case the input wavefront has 0.34 radians rms phase and the
reconstruction residual has 0.1 radians rms. At a wavelength of 4.8$\mu$m 0.1
radians corresponds to 76nm.}
\label{fig:jwst}
\end{figure*}

In general we may expect a combination of segment-base wavefront errors, of
which piston, tilt and focus are correctable. We split these into pure segement
defocus and pure segment piston as a demonstration of correcting single even
functions on the JWST segmented pupil. Both defocus and piston introduce
nonzero fringe phases in the NRM image (since the NRM holes are smaller than
the segments). In reality, piston error on a curved segment contributes other
aberrations. We then mix segment piston, tilt, and focus errors (Fig.
\ref{fig:jwst}-bottom) and can reconstruct both the symmetric and antisymmetric
wavefront error terms unambiguously. While the accuracy of wavefront retrieval
is reduced at segment edges and obstructions, the general structure and the
sign of the phase is preserved in the reconstruction. Fine tuning this
procedure may manage segment edges and obstructions either with different
smoothing function, or by incorporating a gain factor for known
discontinuities.

\subsection{Secondary Mirror Misalignment\label{jwstmimf}}
JWST wavefront sensing has been developed around using NIRCam. NIRCam uses
specialized hardware to determine segment co-phasing, and large and small
wavefront aberrations.  All wavefront sensing is initially performed at one
field point in NIRCam.  This telescope phasing may result in a sub-optimal
positioning of the secondary mirror due to a degenerate combination of certain
primary and secondary misalignment modes that cannot be sensed at a single
field point.  In order improve the wavefront in other JWST instruments,
wavefront sensing needs to be performed in the other imaging instruments,
namely NIRISS and JWST's Mid-Infrared Imager (MIRI).  The most thoroughly
tested approach to this Multi-Instrument Multi-Field (MIMF) wavefront sensing
\cite{mimf} involves sweeping through focus with a secondary mirror move and
using a focus-diverse Miselle-Gerchberg-Saxton phase retrieval algorithm
\cite{GS72,Misell1973}.  Using our approach, we can accomplish an unambiguous
wavefront measurement by combining NIRISS' CLEARP and NRM pupil masks without
invoking a secondary mirror move.  This might enable a shorter commissioning
period for JWST, since the secondary mirror focus sweep is a slow exercise.

\begin{figure*}[htbp!]
\centering
\includegraphics[scale=0.8]{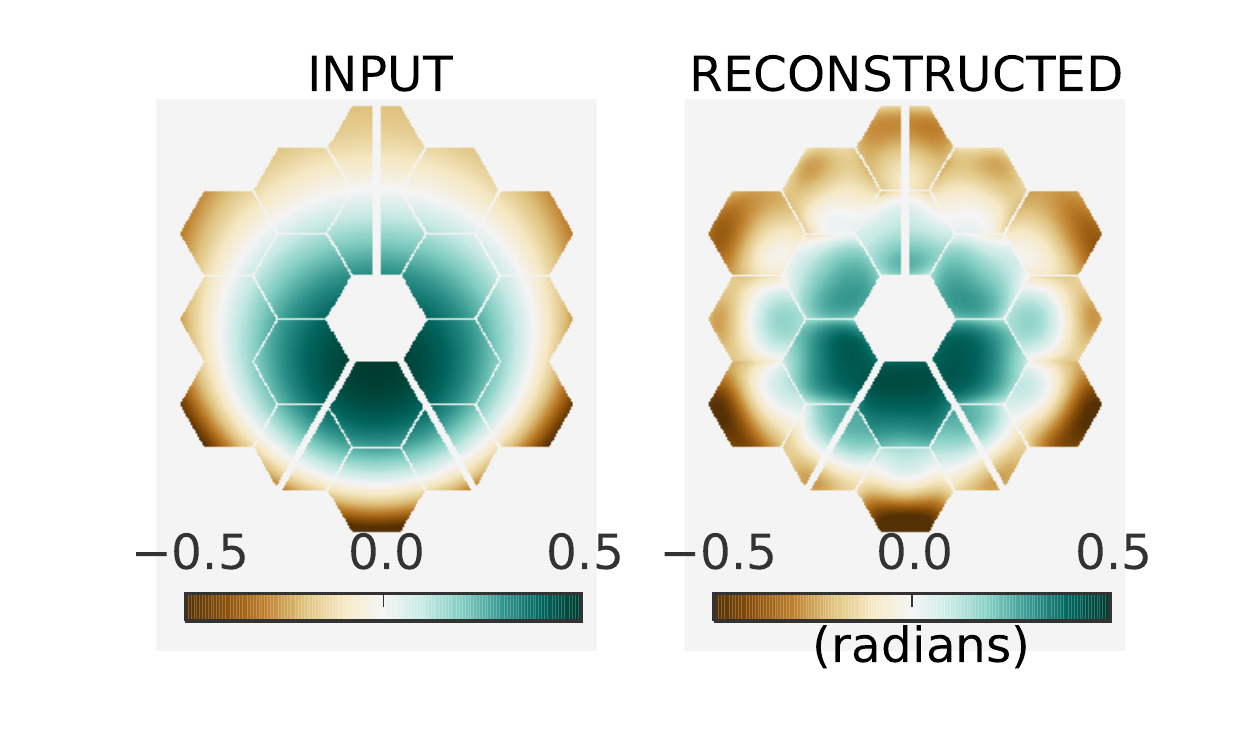}
\caption{Unaberrated segments with secondary mirror misalignment (MIMF)
\cite{mimf}.  Our constrained GS approach will be able to reconstruct low order
global aberrations in the  F480M filter on NIRISS, observing with a star of
magnitude of 7.5 and an exposure time of $<1$s, with $\sim7$~mas jitter.  In
this test case the initial wavefront error is 0.22 radians rms and the
reconstruction residuals are 0.04 radians rms.}
\label{fig:mimfjwst}
\end{figure*}

A mixture of astigmatism and coma are likely to result from the secondary
mirror misalignment.  In NIRISS this aberration might be up to 200 nm in size.
In Fig.~\ref{fig:mimfjwst} we show a simulated measurement of this kind of
aberration using our method. From this single measurement we cannot
determine where the aberrations are from (primary or secondary misalignments),
but NIRISS can add an additional field point compared to NIRCam's wavefront
measurements. Multiple field points are required to determine the secondary
misalignment without moving the secondary.

\section{Discussion \label{sec:discussion}}

Our constrained Gerchberg-Saxton approach to phase retrieval provides an
efficient way to measure wavefront aberrations on current and future space
telescopes using only in-focus images.  The algorithm is suited to both
continuous and segmented/obstructed pupils, though segment obstructions appear
to limit the performance in our simple implementation. We use the first 15
Zernike or Hexike polynomial to smooth the wavefront in the pupil. We smooth
over the full pupil in the case of a continuous pupil, and segment-wise in the
case of the segmented pupil, using the appropriate set of polynomials.
Alternative basis functions could  handle different wavefront errors better. We
ignore the edge effects of thin pupil obstructions. Mitigating the errors
introduced by these effects may take more study. 

In this paper we have discussed our wavefront retrieval approach primarily in
the context of commissioning JWST, although the method could provide wavefront
knowledge during JWST science observations (for example, to support image
deconvolution).  On JWST this work may help avoid some secondary mirror focus
sweeps during the commissioning phase of the telescope. We have presented a
proof of concept of our method;  further optimization of the method can be
tailored to individual cases. 

On JWST, two in-focus exposures containing $10^6$ photons each (requiring
exposure times of $<1$s full pupil images and seconds in NRM images for a star
of $7.5^{\mathrm{th}}$ magnitude through the F480M filter) will provide enough
signal to measure the wavefront errors of $\sim100$ nm, even in the presence of
jitter, finite bandwidth and limited image size. Chromatic smearing, and finite
image size are larger sources of error than the anticipated pointing jitter of
JWST. Our method can tolerate up to 16 mas of pointing jitter, which is twice
as large as JWST's required pointing accuracy. Frequent monitoring of mirror
segment drift can be measured with our approach on NIRISS (which contains the
7-hole NRM used in this study) as a part of normal telescope operations. This
provides a complimentary capability to trend wavefront stability over time in
NIRISS alongside the main wavefront sensing monitoring program using NIRCam.

We have focused on using a non-redundant mask to break the phase
degeneracy in the unconstrained GS algorithm.  It may also be possible to
accomplish this with a redundant pupil that possesses asymmetries, using the
asymmetric pupil wavefront sensor (APWFS) method \cite{Martinache_APWFS}, which
also uses in-focus images. The APWFS algorithm could be used on data from other
JWST instruments, such as NIRCam or MIRI. 

Our method has some key differences compared to the differential optical
transfer function approach to wavefront sensing. The dOTF method similarly
requires two in-focus images, one with a pupil modification, though the
modification must be small. For JWST, Codona \cite{2015JATIS...1b9001C}
suggests using small motions of the pupil wheel to block a portion of the pupil
and achieve its required pupil diversity, which is not possible for MIRI's
rachet-mechanism filter wheel. Our constrained GS approach uses two standard
filter settings on NIRISS. With the possibility of using APWFS measurements, a
single image could suffice for doing constrained GS wavefront sensing with
MIRI.

Using these methods for both monolithic and segmented future telescopes
(such as the Wide Field Infrared Survey Telescope \cite{WFIRST} or the
High Definition Space Telescope \cite{HDST})  can utilize science hardware for
wavefront sensing, which has obvious benefits for weight, cost, complexity, and
scope.

\section{Acknowledgments}

We thank Marshall Perrin and Joseph Long for helpful discussions and support
for WebbPSF software. We also thank an anonymous referee for helpful and
scholarly comments.  This work was supported by the National Science Foundation
Graduate Research Fellow Grant No. DGE-123825, NASA APRA Grant NNX11AF74G, and
the STScI Director's Discretionary Research Fund. This research made use of
Astropy, a community-developed core Python package for Astronomy
\cite{2013A&A...558A..33A}.

\end{document}